\documentclass[useAMS,usenatbib]{mn2e}  
\usepackage{graphicx}
\title{Modelling the $\gamma$-ray variability of 3C 273}
\author[Y.G. Zheng et al.]{Y.G. Zheng$^{1}$, L. Zhang$^{2}$\thanks{Corresponding author. E-mail: lizhang@ynu.edu.cn}, B.R. Huang$^{1}$, S.J. Kang$^{3}$\\
       $^{1}$Department of Physics, Yunnan Normal University, Kunming, Yunnan, China\\
       $^{2}$Department of Physics, Yunnan University, Kunming, Yunnan, China\\
       $^{3}$School of Physics, Huazhong University of Science and Technology, Wuhan,China}

\begin{document}
\date{Received date  / Accepted date}

\pagerange{\pageref{firstpage}--\pageref{lastpage}} \pubyear{2006}

\maketitle

\label{firstpage}

\begin{abstract}
We investigate MeV-GeV $\gamma$-ray outbursts in 3C 273 in the frame of a time-dependent one-zone synchrotron self-Compton (SSC) model. In this model, electrons are accelerated to extra-relativistic energy through the stochastic particle acceleration and evolve with the time, nonthermal photons are produced by both synchrotron and inverse Compton scattering of synchrotron photons. Moreover, nonthermal photons during a quiescent are produced by the relativistic electrons in the steady state and those during a outburst are produced by the electrons whose injection rate is changed at some time interval. We apply the model to two exceptionally luminous $\gamma$-ray outbursts observed by the Fermi-LAT from 3C 273 in September, 2009 and obtain the multi-wavelength spectra during the quiescent and during the outburst states, respectively. Our results show that the time-dependent properties of outbursts can be reproduced by adopting the appropriate injection rate function of the electron population.
\end{abstract}


\section{Introduction}
\label{sec:intro}

Blazars, a special class of active galactic nuclei (AGN), exhibit that the continuum emission, which arise from the jet emission taking place in AGN whose jet axis is closely aligned with the observer's line of the sight, is dominated by nonthermal emission as well as rapid and large amplitude variability. The broad spectral energy distributions (SED) from the radio to the $\gamma$-ray bands is dominated by two components, indicating two humps. It is widely admitted that the first hump is produced from electron synchrotron radiation, the peaks range from the infrared-optical up to X-ray regime for different blazars (Urry 1998). The second hump, the peaks in the GeV to TeV  $\gamma$-ray band, probably is produced from inverse Compton scattering of the relativistic electrons either on the synchrotron photons (Maraschi et al. 1992) or on some other photon populations (Dermer et al. 1992; Sikora et al. 1994).

Flaring activity is one of the important features of AGN. Particularly in the X-ray and TeV regimes, nonthermal photons are produced by radiation of ultra-relativistic electrons close to their maximum energy. Recently, following the development of observational techniques in the $\gamma$-ray range, some extreme and remarkable variability behaviors have been caught by ground-based Cherenkov telescope experiments and space telescope experiments, for example, HESS for PKS 2155-304 (Aharonian et al. 2007), MAGIC for Markarain 501 (Albert et al. 2007), VERITAS for 1ES 1218+304 (Acciari et al. 2009) and 3C 66A (Acciari et al. 2009), Fermi-LAT for 3C 273 (Abdo et al. 2010). In order to explain for above variability behaviors, a wealth of $\gamma$-ray emission models have been developed (Mastichiadis \& Moraitis 2008; Bednarek \& Wagner 2008; Weidinger \& Spanier 2010; R$\ddot{u}$ger et al. 2010; Zheng \& Zhang 2011).

As the brightest and nearest ($z = 0.158$) quasar, 3C 273 is significant for studying AGN. Studies of this source are crucial to understand the AGN physics and to put reliable constraints on jet parameters. In order to do so, in the recent years, some intensive observational campaigns were carried out in the X-ray bands (e.g. Kataoka et al. 2002; Courvoisier et a. 2003; Grandi et al. 2003). Observational X-ray features imply that 3C 273 is a unique object whose hard X-ray emission occasionally contains a component that is not related to a beamed emission, but likely to originate in inverse Compton scatter from the relativistic jet(Kataoka et al. 2002). Grandi et al. (2003) analyzed a long-term X-ray spectral variability of 3C 273, and untangled the jet and accretion disk emission in the object. Though, In the $\gamma$-ray bands, it was observed by EGRET ten years ago(Hartman et al. 1999), the source did not show a significant result. In September 2009, two exceptionally luminous $\gamma$-ray outbursts are observed by Fermi-LAT (Abdo et al. 2010). This allowed us to study in their spectral and temporal structures. In this paper, we apply the time-dependent one-zone synchrotron self-Compton (SSC) model (e.g. Katarzynski et al. 2006; Zheng \& Zhang 2011)to 3C 273 for explaining its flare property, especially two exceptionally luminous $\gamma$-ray outbursts in September 2009. Our goal is to determine whether the model can give reasonable fitting parameters for both in the stable (pre-burst) and variable (in-burst) states, and reproduce two significant outbursts in a month.

Throughout the paper, we assume the Hubble constant $H_{0}=70$ km s$^{-1}$ Mpc$^{-1}$, the matter energy density $\Omega_{\rm M}=0.27$, the radiation energy density$\Omega_{\rm r}=0$, and the dimensionless cosmological constant
$\Omega_{\Lambda}=0.73$.

\section{Model}
\label{sec:model}

We basically follow the approach of Zheng \& Zhang (2011) to model the temporal evolution of photon emission from the 3C 273 through two parts: the temporal evolution of particle energy distributions and production of photons. Below, we give a brief description on the model (Please see Katarzynski et al. (2006), Zheng \& Zhang (2011) or Zheng et al. (2011) for more details about the model).

We start with the one dimensional momentum diffusion equation (Tverskoi 1967; Schlickeiser 1984), where the relativistic approximation of the dimensionless particle momentum $p\approx \gamma$ is used and $\gamma$ is the particle Lorentz factor. The evolution of the energetic particle distribution can be written as
\begin{eqnarray}
\label{Eq:1} \frac{\partial N(\gamma,t)}{\partial
t}=\frac{\partial}{\partial
\gamma}\{[C(\gamma,t)-A(\gamma,t)]N(\gamma,t)+ D(\gamma,t)\frac{\partial
N(\gamma,t)}{\partial
\gamma}\}+Q(\gamma,t)-E(\gamma,t)\;,
\end{eqnarray}
where $C(\gamma,t)=\frac{4\sigma_{\rm T}c}{3m_{\rm e}c^2}[U_{\rm B}(t)+U_{\rm rad}(\gamma, t)F_{KN}]\gamma^2$ is the radiative cooling parameter that describes the synchrotron and inverse-Compton cooling of the particles at time $t$, where $t$ is in the unite of acceleration timescale, $m_{\rm e}$ is the electron rest mass, $\sigma_{\rm T}$ is the Thomson cross section, $U_{\rm B}$ and $U_{\rm rad}$ are magnetic and radiation field energy densities respectively, $F_{\rm KN}$ is corrected coefficient of the Klein-Nishina (KN) effects (e.g. see Moderski et al. 2005); $A(\gamma, t)=\gamma/t_{\rm acc}$ is the acceleration term that describes the particle energy gain per unit time and $t_{\rm acc}=\gamma^2/2D(\gamma, t)$ is the acceleration time; $E(\gamma,t)=N(\gamma, t)/t_{\rm esc}$ represents escape term and $t_{\rm esc}=R/c$ is escape timescale that depends on the emission region size $R$; and $Q(\gamma,t)$ is the source term, here we consider continuous injection case, i.e. the particles are continuously injected at the lower energy ($1\le \gamma\le2$) and systematically accelerated up to the equilibrium energy.

In order to obtain the distribution function of electrons, this time-dependent kinetic equation is solved in a spherical and homogeneous zone in which contains isotropically distributed electrons and a randomly oriented magnetic field. We adopt an implicit difference scheme given by Chang \& Copper (1970). After calculating the electron number density $N(\gamma, t)$ at a time $t$, we can calculate the synchrotron intensity $I_{\rm s}(\nu, t)$ and the intensity of self-Compton radiation $I_{\rm c}(\nu, t)$, and then calculate the flux density observed at the observer's frame from the radiation intensity.

In the TeV energy ranges, the extragalactic background light (EBL) absorption effect is significant (Kneiske et al. 2004; Dwek \& Krennrich 2005; Franceschini 2008; Kneiske 2010). In our calculations, because of the MeV-GeV energies two orders of magnitude under the energy range that effected by EBL absorption, we need not take into account the EBL absorption.

\section{Apply to 3C 273}
\label{sec:apply}

3C 273 is a typical flat spectrum radio quasar (FSRQ). Although abundance of data of 3C 273 are collected and the time structures are investigated by Soldi et al. (2008), the surprising phenomenon still brings on the new argue for physical model. Below, we will use our model to the outburst events observed by Fermi-LAT in September 2009 in 3C 273 (Abdo et al. 2010). The spectra in the stable (pre-burst) and variable (in-burst) states are divided by the time. During the quiescent phase preceding the September events, we computed the spectra integrating over the two weeks from September 1 to 13, and the spectra of the two major outbursts were integrated over the time intervals from September 14 to 18 and from September 19 to 23 (Abdo et al. 2010).

\subsection{Photon Spectrum in Pre-Burst State}

We apply the model described in \S 2 to 3C 273 and model its $\gamma$-ray variability. In order to do so, following assumptions are made: (1) A constant initial electron distribution $N_{ini}(\gamma, 0)=7.0\times10^{-3}$ cm $^{-3}$ for $1\leq\gamma\leq2$ and the injection rate of the electron population is $Q(\gamma)=7.0\times10^{-3}$ cm$^{-3}$ s$^{-1}$ for $1\leq\gamma\leq2$, (2) acceleration timescale is equal to constant escape timescale during the evolution process, i.e. $t_{\rm acc}=t_{\rm esc}=R/c$, and (3) minimum and maximum Lorentz factors of electrons are  $\gamma_{\rm min}=1$, $\gamma_{\rm max}=10^{5}$, magnetic field strength is $B=0.97$ G, emission region size is $R=1.0\times10^{16}$ cm, and Doppler factor is $\delta=12$. Under above assumptions, we numerically solve Eq. (1) following Zheng \& Zhang (2011) and find that the electron's evolution tends to the steady state when $t=15t_{\rm acc}$. Since relativistic electrons should be in the steady state for the gamma-ray emission in the quiescent (pre-burst) state, we can calculate the electron's evolution spectrum trends to the steady state spectrum and then calculate the quiescent $\gamma$-ray spectrum in the one-zone SSC model in \S 2.

\begin{figure}
\centering
\includegraphics[scale=1.0]{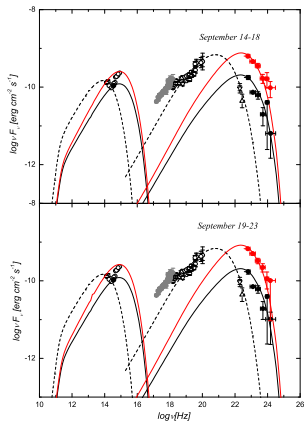}
\caption{Comparisons of predicted multi-wavelength spectra with observed data of 3C 273 in
the September, 2009, and in the history. The black solid curves and the red solid curves represent the SSC model for the quiescent state and the outburst state spectrum, respectively. The SSC model for the other state spectrum is reported for comparison as dashed curves. Top panel shows the outburst on the September 14-18, 2009, and bottom panel shows the outburst on the September 19-23, 2009. The high energy observed data come from Abdo et al. (2010), and the historical data come from Pacciani et al. (2009). }
\label{Fig:1}

\end{figure}

In Fig. \ref{Fig:1}, we show predicted the quiescent spectrum from radio to TeV $\gamma$-ray bands (black solid curves) together with the predicted other state spectrum (dashed curves), that represent the multi-band photon spectrum during the pre-burst in September 2009 and in the history, respectively. For comparison, observed data of 3C 273 at MeV-GeV bands from September 1 to 13, 2009 (Abdo et al. 2010) and the multi-wavelength historical data (Pacciani et al. 2009)are also shown, where black solid circles with error bars represent the observed values during from September 1 to 13, 2009 and open symbols represent the historical values. It can be seen that the observed data in the quiescent state and in the other state can be reproduced in the model.

\begin{table*}
\begin{minipage}[t][]{\textwidth}
\caption{Physical parameters of the time-dependent one-zone SSC model spectra}
\label{table1}
\begin{tabular}{lccc}
\hline\hline

parameters & history data & pre-burst state & outburst state \\

$\rm \gamma_{min}$   & 1  & 1 & 1 \\
$\rm \gamma_{max}$ & $10^{5}$ & $10^{5}$& $10^{5}$ \\
$\rm R~[cm]$   & $3.55\times10^{16}$ & $1.0\times10^{16}$ & $1.0\times10^{16}$ \\
$\rm B~[G]$ & 0.71 & 0.97 & 0.97 \\
$\rm \delta$   & 5.5 & 12 & 12 \\
$\rm Q~[cm^{-3}~s^{-1}]$   & $3.8\times10^{-1}$ & $7.0\times10^{-3}$ & injection rate function \\
$\rm N_{ini}~[cm^{-3}]$   & $3.8\times10^{-1}$ & $7.0\times10^{-3}$ & steady-state electron spectrum \\
\hline\\
\end{tabular}
\end{minipage}
\end{table*}

\subsection{Photon Spectrum in Outburst State}

We now consider the case in the outburst state. We assume that some fresh electrons are injected during the outburst state, i.e. the change of the photon spectrum in the outburst state with respective to that in the pre-burst state is due to the change of electron's injection rate at some time intervals. In other words, model parameters in the outburst state are the same as those in the pre-burst state except for the electron's injection rate. For 3C 273, in order to reproduce the observed MeV-GeV photon spectrum over the time intervals from September 14 to 18 and from September 19 to 23, and reproduce the light curves in September 2009 in a same time series, we adopt the injection rate function of the electron population
\begin{eqnarray}
Q(\gamma,t)=Q_{\rm
ini}H(2-\gamma)[6.7\Theta(0.5-t)+7.2\Theta(t-12.5)\Theta(13-t)]\;,
\label{eq:Fr}
\end{eqnarray}
where $Q_{\rm ini}=7.0\times10^{-3}$ cm$^{-3}$ s$^{-1}$ is injection rate constant, $H(\gamma)$ and $\Theta(t)$ is the Heaviside function, respectively, and $t$ is evolution time normalized to the acceleration timescale.

Using the model parameters which are same as those in the pre-burst state and consider the resulting steady-state electron spectrum as an initial condition, we numerically solve Eq. (1) with the injection rate of Eq. (2) and reproduce the observed MeV-GeV photon spectrum (red solid curves) of 3C 273 over the time intervals from September 14 to 18 and from September 19 to 22, 2009 in Fig. 1. For comparison, observed data of 3C 273 at MeV-GeV bands from September 14 to 18, and from September 19 to 23, 2009 (Abdo et al. 2010) are also shown as red solid circles with error bars. It can be seen that the observed data in the outburst state can also be reproduced in the model.

\subsection{Energy Dependent Light Curves}

\begin{figure}
\centering
\includegraphics[scale=.7]{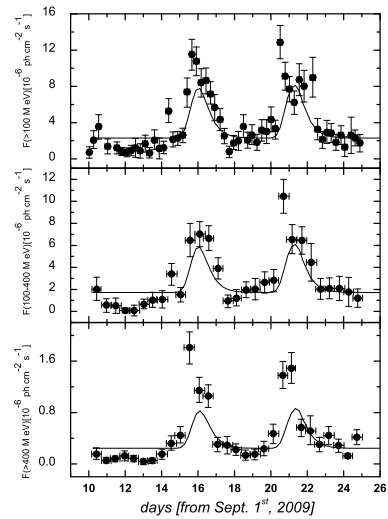} \caption{Comparisons of
simulated light curves (solid lines) with observational light curve
(data points) from the Abdo et al. (2010) for the September,
2009. Two exceptionally luminous $\gamma$-ray outbursts can be seen in the simulated light curves.}
  \label{Fig:2}
\end{figure}

After numerically solving Eq. (1) with the injection rate of Eq. (2) and  the model parameters which are same as those in the pre-burst state, we calculated energy-dependent light curves at energy bands of $>$100 MeV, 100-400 MeV, $>$400 MeV, respectively, where the integrated fluxes are estimated by using the differential spectra showed in Fig. \ref{Fig:1}. Our results are shown in Fig. \ref{Fig:2}. For comparison, the observational light curves of 3C 273 in September 2009 are also shown in Fig. \ref{Fig:2}. The timescales are transformed as the time in the observer's frame. It can be seen from Fig. \ref{Fig:2} that our model can reproduce the outbursts at the energy bands of $>$100 MeV, 100-400 MeV, and $>$ 400 MeV.

\subsection{Variability profile}

\begin{figure}
\centering
\includegraphics[scale=.3]{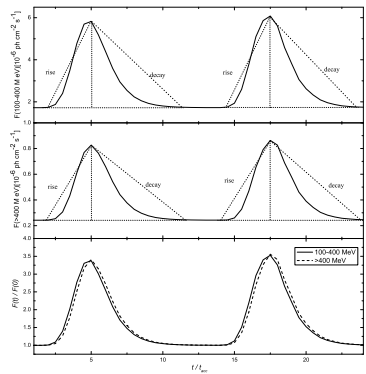} \caption{Predicted light curves(solid lines) variability profile at 100-400 MeV(top panel), and $>$400 MeV (middle panel). The first peak shows the flare from September 14 to 18, and the second peak shows the flare from September 19 to 23. The dotted line shows the variability profile. A clearly asymmetric triangle variability profile can be seen. In the bottom panel, we give the time lag analysis results for the two exceptionally luminous $\gamma$-ray outburst of 3C 273 in September 2009. The fluxes at different wavelengths are normalized to the pre-burst state value. Our numerical results indicate that there is no visible time lag of the peak in the light curves.}
  \label{Fig:3}
\end{figure}

\begin{figure}
\centering
\includegraphics[scale=.5]{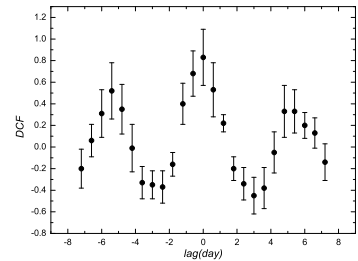}
\caption{The observed time lag between 100-400 MeV and $>$ 400 MeV with the discrete correlation function. As can be seen that the observed light curves do not show a significant time lag.}
\label{Fig:4}

\end{figure}

In order to discern the profile of the flare, in Fig. \ref{Fig:3}, we decompose the predicted light curves at MeV-GeV bands from September 14 to 18, and from September 19 to 23, 2009. It can be seen from Fig. \ref{Fig:3} that the flares at different energies can be described by a asymmetric triangle variability profile. The significant events have light curves characterized by fast rise of the duration tomes of about 3.1$t_{\rm acc}$, and a longer decay of about 6.3$t_{\rm acc}$, These timescales correspond to about 1 day and 2.5 days in the Earth's frame, respectively. Radiation cooling induces to a long tail in the light curves. Our results indicated that the light curves do not show the visible time lag of the peaks in the light curves between 100 - 400 MeV and $>$ 400 MeV. In Fig. \ref{Fig:4}, we show the observed time lag between 100-400 MeV and $>$ 400 MeV with the discrete correlation function (DCF; Edelson \& Krolik 1988; Zheng et al. 2008). As can be seen that observed light curves do not show a significant time lag. These are in agreement with our numerical results.

\begin{figure}
\centering
\includegraphics[scale=.3]{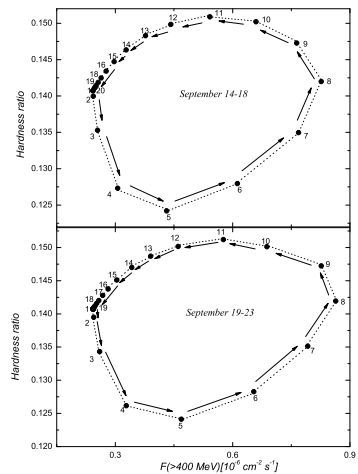} \caption{Simulated hardness
ratio $\frac{F(100 - 400 \rm MeV)}{F(> 400 \rm MeV)}$ vs. flux
F($>$400 MeV) for the outbursts from September 14 to 18 (top
panel), and from September 19 to 23 (bottom panel), 2009. Black
circles represent flaring emission. The number near the markers
denote the position of the points in the light curves. A canonical
counterclockwise loop is seen.}
  \label{Fig:5}
\end{figure}

We also calculate the evolution of the hardness ratio which is defined as the ratio $\frac{F(100 - 400 \rm MeV)}{F(> 400 \rm MeV)}$. The evolution of the hardness ratio with the emitted flux above 400 MeV is shown in Fig. \ref{Fig:5}. It can be seen from Fig. \ref{Fig:5} that the evolution of the flare points show a clear canonical counterclockwise loop.

\section{Discussion}
\label{sec:discussion}

In this paper, we have examined the expected MeV-GeV variability of 3C 273 in the context of the time-dependent one-zone SSC model. By varying the injection rate parameter of the model, we found that the outburst events of September 2009 are induced. Our calculation shows that a one-zone SSC model incorporating time evolution of the electron and photon distributions can be a viable tool to study the rapid variability of blazars. Here, we focused on the stochastic particle acceleration origin of the variability. Obviously, variability can also be triggered by a change in other model parameters (e.g. Kirk et al. 1998; Kataoka et al. 2000; Moraitis \& Mastichiadis 2010). The main idea of our method is to decompose the quasar spectrum in the low luminosity (quiescent) state and high luminosity (outburst) state, the spectrum in the high luminosity state is produced by injection fresh electron mixing with the electron distribution in the low luminosity state. Despite its limitations, the model can reproduce two significant
outbursts at high energy bands. The 100 MeV-400 MeV luminosity exhibits almost simultaneous variations with $>$ 400 MeV luminosity. Though the model can also predict variabilities at other wavelengths, unfortunately, in September 2009, the apparent position of 3C 273 is close to the Sun, it is impossible to obtain other wavelength light curves.

We adopt the physical parameters for the model give a good representation of the observational data. Lister et al. (2009) makes a detail plot of the kinematics of the various components in the radio jet of 3C 273, this result mean superluminal velocity is $\rm \beta_{app}=13.4$, with an estimated Doppler factor $\rm \delta=16.8$ (Abdo et al. 2010). To describe the 3C 273 flare in September 2009, in our calculation, we adopt the Doppler factor $\rm \delta=12$, this value is in agreement with the kinematical estimation. As for the emission region size, we can self-consistently determined from the variability timescale by $\rm R\leq c\Delta t_{obs}\delta$. If the variability timescale of the source on the a day order, we can give the emission region size $\rm R\sim 2.5\times10^{15}\delta$ cm. Furthermore, we adopt the shorter acceleration and escape timescales with $t_{\rm acc}=t_{\rm esc}=\frac{R}{c}=t_{\rm cr}$ than other investigators (Kirk et al. 1998; Mastichiadis \& Moraitis 2008). These assumptions can lead to higher acceleration rate and lower escape rate, and make more particles acceleration up to high energy rapidly. In order to reproduce both MeV-GeV radiation and variability of 3C 273, we change the injection rate of the low energy particles. It should be noted that when the shock front overruns a region in the jet in which the local plasma density is enhanced. The number of particles increase as an avalanche occurring in the jet, the injection rate can be expected to change (e.g. Zheng \& Zhang 2011).

Our results show that the outbursts are triggered by a magnetized cloud in a relatively small region, in which the stronger magnetic field and larger bulk factor with a low constant electron number density exist. Stochastic acceleration fresh particles induced the synchrotron and inverse-Compton peak shift toward higher energies. The shifts of the inverse-Compton peak from observation to observation were previously proposed from the comparison of the June 1991 multi-wavelength campaign and the OSSE observation of September 1994 (McNaron-Brown et al. 1997). Based on the multi-frequency observation and modelling results, Pacciani et al. (2009) argue that this behaviour is a more general feature of the source, happening on shorter timescales. Given the complexity of the energy spectra variability of the source, this requires more detailed observations and that the issue remains open.

It can be seen that the calculated light curves show the character with the decay time longer than the rise time. Generally, the radiative cooling of high energy electrons can induce the symmetric variability profile. In the model, we have contained the acceleration, cooling, injection, and escape processes. Kataoka et al. (1999) argue that the rapid variability events observed in blazars are well characterized by four dynamical timescales, i.e. acceleration time: $t_{\rm acc}$, cooling time: $t_{\rm cool}$, source light travel time: $t_{\rm crs}$, and electron injection time: $t_{\rm inj}$. If $t_{\rm cool}>t_{\rm acc}\sim t_{\rm inj}\sim t_{\rm crs}$, the triangle asymmetric variability outline can be expected (Kataoka et al. 1999). In order to reproduce both high energy radiation and variability of 3C 273, we assumed the $t_{\rm acc}=t_{\rm crs}=R/c$, $t_{\rm inj}=0.5t_{\rm acc}$. Using the parameters in \S 3, we can calculate that the equilibrium energy at $t_{\rm acc}=t_{\rm cool}$ is $\gamma_e\approx 2.5\times10^{3}$. Since the synchrotron radiation are most produced near the equilibrium energy $\gamma_{e}$, we can expect that the up-scatter soft photons energy are located on the $\sim 4.13$ eV. Then the electron energy which emit MeV-GeV $\gamma$-ray photons by inverse Compton scattering is $\gamma\approx 0.3\times10^{4}$. This energy is comparable with the equilibrium energy. Tammi \& Duffy (2009) suggested that the electrons whose energy are less than the equilibrium energy are dominated by acceleration process with $t_{\rm cool}>t_{\rm acc}$, so we can conclude that the $t_{\rm cool}(\gamma)\sim t_{\rm acc}$ for MeV-GeV $\gamma$-ray radiation, which probably lead to the asymmetric variability profile in the light curves.

This is more clearly seen that variability patterns track a loop in opposite direction in the 100 MeV-400 MeV and $>$ 400 MeV energy bands, characterized with anti-clockwise motion.  The hardness ratio change will result in some Compton peak moving (e.g. Albert et al. 2007). The behavior of the mean multi-frequencies spectra during the pre-burst and during the outbursts is a little different. During the flare, the peaks of both synchrotron and IC emissions move to lower frequencies. We argue that this can be explained by the energy loss of the electrons during the outbursts. The possible patterns in clockwise hysteresis were discussed by Kirk et al. (1998). They argued that clockwise pattern arises whenever the spectral slope is controlled by synchrotron cooling or any cooling process which is faster at higher energy so that information about changes in the injection propagates from high to low energies (Tashiro et al. 1995), while anti-clockwise pattern arises whenever the spectral slope is controlled by acceleration process. In this case, information about the occurrence of a flare propagates from lower to higher energy, as particles are gradually accelerated into the radiating window. The anti-clockwise loop suggests that the most natural interpretation of the flaring events in 3C 273 is that they are due to the acceleration of the lower energy electrons responsible for the $\gamma$-ray emission.

The soft lags where lower energy radiation peaks later than the higher energy one are easier to explain by high energy radiating particle cooling and radiating on lower and lower energy, but, intuitively, the hard lags would require particles not cooled but heated or accelerated during the flare. In this view, the stochastic particle acceleration origin of the variability would induce to the hard lags. Our results indicated that the light curves do not show the time lag of the peaks in the light curves between 100 - 400 MeV and $>$ 400 MeV. We argue that the timescales that accelerate the particles energy up to $\gamma=0.3\times10^4$ are comparable with the timescales that accelerate the particles energy up to $\gamma=1.2\times10^4$ (see, e.g. Tamma \& Duffy 2009). The longer observational time bins than the acceleration timescales probably lead to the radiative lifetimes practically loss.

\section*{Acknowledgments}
We thank the anonymous referee for valuable comments and suggestions. This work is partially supported by the National Natural Science Foundation of China under grants 11178019, U1231203 and the Natural Science Foundation of Yunnan Province under grants 2011FB041. This work is also supported by the Science Foundation of Yunnan educational department (grant 2012Z016).


\end{document}